\documentclass{article}

\usepackage{microtype}
\usepackage{graphicx}
\usepackage{subcaption}
\usepackage{booktabs} % for professional tables

\usepackage{hyperref}

\usepackage[preprint]{icml2026}

\usepackage{amsmath}
\usepackage{amssymb}
\usepackage{mathtools}
\usepackage{amsthm}
\usepackage{makecell}

\usepackage{xcolor}
\usepackage[normalem]{ulem}

\usepackage[capitalize,noabbrev]{cleveref}

\theoremstyle{plain}

\theoremstyle{definition}

\theoremstyle{remark}

\icmltitlerunning{PanoMHD: Multimodal Modelling of Plasma Dynamics towards Tokamak Control}

\begin{document}

\twocolumn[
  \icmltitle{PanoMHD: Multimodal Modelling of Plasma Dynamics towards Tokamak Control}
  \icmlsetsymbol{intern}{*}

  \begin{icmlauthorlist}
    \icmlauthor{Hyeongjun Noh}{snu,intern}
    \icmlauthor{Chweeho Heo}{snu}
    \icmlauthor{Xiaotian Gao}{zgc,msra}
    \icmlauthor{Yong-Su Na}{snu}
  \end{icmlauthorlist}

  \icmlaffiliation{snu}{Seoul National University, Republic of Korea}
  \icmlaffiliation{zgc}{Zhongguancun Academy, China}
  \icmlaffiliation{msra}{Microsoft Research Asia, China}

  \icmlcorrespondingauthor{Xiaotian Gao}{gaoxiaotian@bza.edu.cn}
  \icmlcorrespondingauthor{Yong-Su Na}{ysna@snu.ac.kr}

  \icmlkeywords{Nuclear Fusion, Plasma Physics, Multimodal, Applied Machine Learning}

  \vskip 0.3in
]

\printAffiliationsAndNotice{\textsuperscript{*}Work done during an internship at Microsoft Research.} 

\begin{abstract} 
Modelling the dynamics of complex physical systems is a fundamental challenge, particularly where nonlinear dynamics and multi-scale interactions render traditional simulations computationally prohibitive. Nuclear fusion plasma represents a complex system where accurately predicting the plasma state, encompassing both performance and stability, is a prerequisite for active control required for sustained energy production. However, existing approaches are limited in providing a comprehensive solution as they largely focus on predicting isolated indicators such as binary stability labels. To overcome this, we present Panoramic MagnetoHydroDynamics (PanoMHD), a self-supervised multimodal framework designed to model plasma dynamics. By utilising a causal Transformer operating on tokenised representations of multimodal physical signals, PanoMHD is able to model the dynamics of high-dimensional magnetic fluctuation signals, which serve as a direct signature of plasma stability. This shifts the prediction paradigm from isolated indicators to multimodal signals. We pioneer the direct prediction of magnetic fluctuation signals for the first time, and demonstrate that this comprehensive representation enables state-of-the-art performance on KSTAR nuclear fusion plant experimental data. Our model outperforms baselines in future plasma performance prediction ($R^2=0.987$ vs. $0.957$) and surpasses dedicated classifiers in the downstream classification of distinct plasma states (L/H mode) with 97.3\% vs. 94.5\% accuracy.
\end{abstract}

\section{Introduction}

Nuclear fusion has emerged as a promising next‑generation baseload energy source, with zero carbon emissions, intrinsic safety, and a limitless fuel supply. Among various approaches to achieve fusion, the tokamak has demonstrated the most advanced experimental progress over the past decades, capable of confining a plasma of several hundred million degrees for more than tens of seconds in a doughnut-shaped toroidal vacuum vessel using strong magnetic fields.

The fundamental objective of a tokamak is to sustain fusion plasmas at sufficiently high temperature and density with adequate energy confinement time \cite{JDLawson_PPSSB_1957}. Although experiments have demonstrated significant progress towards this goal \cite{Strachan_PRL, LaoLL_PRL, Koide_PRL, Luce_2014, Hybrid, FIRE, Na_2025}, substantial challenges remain. A primary challenge is controlling the plasma at higher performance (i.e., higher temperature, density, and energy confinement time) while maintaining stability \cite{HZohm_PPCF_2003}.This requires an accurate dynamics model for fusion plasma to predict performance and instabilities.

For modelling plasma dynamics, various physics-based and data-driven approaches have been investigated. However, the tokamak plasma is a highly complex physical system, making this task inherently challenging. While physics-based frameworks \cite{CLee_NF_2021, Brizard_RevModPhys_2007} have been employed, they face a critical trade-off between fidelity and computational cost. Reduced models can be efficient but may not fully reproduce experimental behaviour, whereas high-fidelity approaches are often too expensive for real-time simulations. To overcome these numerical limitations, artificial intelligence (AI) has recently been incorporated \cite{TGLFNN, QuaLiKizNN, joung_gs-deepnet_2023, TORAX, SEO20245396, Heo_2024, HJNoh_JCP_2025, Chung_2025}, but the reliable modelling of such coupled systems remains far from solved.

On the other hand, given the difficulties in physics-based plasma modelling, many data-driven approaches have been developed for directly predicting 0D plasma performance and confinement-threatening instabilities based on diagnostics (i.e., sensor measurements). In particular, research has particularly focused on predicting disruptions and tearing instabilities \cite{PCVries_NF_2011}, as these destabilising events pose the primary threat of device damage. Consequently, disruption prediction across several tokamaks has been studied \cite{HKHarbeck_Nature_2019, JVega_NaturePhyscis_2022, Orozco_TPS_2022, Kim2024KSTAR, Kim_2024_BNN_KSTAR, LEE2024114128, Ai2024AnomalyKSTAR, Lee_2025}, alongside research on mitigating damage during disruptions \cite{Lehnen2015ITER, Fu2020MLControl} and avoiding tearing instabilities \cite{Seo2023IJCNNTearing, Seo2024NatureRL, sonker2025ICML} have progressed. However, while these studies have advanced plasma stability control, they are typically specialised for specific plasma states or instabilities, necessitating the aggregation of multiple disparate predictors. This fragmentation increases system complexity and compromises end-to-end reliability. Furthermore, these models often rely on dense diagnostic arrays to maximise accuracy, including costly and availability-limited measurements like Thomson scattering, thereby escalating the hardware and integration burdens of real-time control. Finally, these methods are predominantly supervised, requiring extensive labelling and curation, which imposes a heavy workload and hinders scalability and transferability to related tasks, new operating regimes, or other devices.

To address these limitations, we propose Panoramic MagnetoHydroDynamics (PanoMHD), a self-supervised multimodal modelling framework for fusion plasma dynamics. Instead of relying on comprehensive diagnostics to predict isolated instability labels, PanoMHD directly models the multimodal dynamics of fundamental 0D plasma parameters and raw magnetic fluctuations captured by Mirnov coils (MCs), where the latter serve as a direct and physically interpretable signature of instability. Crucially, we restrict our input sources to these MCs and fundamental parameters, utilising only the most reliable, widely available, and cost-effective diagnostics for tokamak control. The resulting predictions provide a solid foundation that can be readily leveraged for a wide range of downstream tasks, effectively reducing the need for multiple specialised predictors and scarce diagnostic resources. In summary, our main contributions are threefold:

\begin{itemize}
    \item \textbf{A robust and cost-efficient solution for tokamak control:} We demonstrate that plasma dynamics can be modelled using only Mirnov coils and standard control commands, bypassing complex diagnostics. This Mirnov-centric approach is vital for future commercial reactors, where high neutron fluxes preclude delicate instrumentation.

    \item \textbf{Pioneering self-supervised multimodal modelling:} To our knowledge, we introduce the first self-supervised framework integrating 0D parameters with high-dimensional magnetic fluctuation spectrograms. Unlike approaches limited to isolated indicators, this multimodal strategy captures comprehensive spectral details of instabilities without explicit event labels.

    \item \textbf{General-purpose foundation with SOTA performance:} We demonstrate PanoMHD as a general-purpose foundation for diverse tasks. It successfully predicts instabilities (e.g., ELMs, tearing instabilities) and achieves SOTA in future plasma performance ($\beta_N$) prediction ($R^2=0.987$ vs. $0.957$). Notably, the predicted magnetic fluctuation spectrograms enable superior accuracy in downstream plasma confinement mode classification (L-mode vs. H-mode), surpassing a dedicated baseline (97.3\% vs. 94.5\%).
    
\end{itemize}

\section{Problem Statement}

We formulate the multimodal modelling of tokamak dynamics as learning a parameterised model $P_\theta$ that predicts the future plasma state. The learning objective is to maximise the log-likelihood of the future magnetic fluctuations $MC_{t+1}$ and plasma performance $p_{t+1}$ conditioned on the history of plasma states ($MC, p$) and control commands ($c$) within a context window of length $L$. The detailed architecture modelling the transition probability $P_\theta(MC_{t+1}, p_{t+1} \mid MC_{t-L+1:t}, p_{t-L+1:t}, c_{t-L+1:t})$ is presented in \cref{Model Architecture of PanoMHD}.

We utilise the KSTAR dataset $\mathcal{D} = \{ \tau^{(i)} \}_{i=1}^N$ described in \cref{KSTAR data}, where each trajectory $\tau^{(i)}$ consists of a sequence of tuples $(MC_t, p_t, c_t)$ discretised at $\Delta t$. Specifically, the magnetic fluctuations $MC_t$ are derived from the MC diagnostics detailed in \cref{Mirnov Coil (MC)}. The subsequent sections outline the remaining components: the tokenisation strategy for these multimodal physical signals is presented in \cref{Data Preprocessing}, and the metrics used to validate the model's performance are defined in \cref{evaluation}.

\section{Methodology}

\subsection{KSTAR data}
\label{KSTAR data}
The dataset comprises $N=978$ shots, or experimental records, from the 2017--2022 KSTAR campaigns \cite{Lee2000KSTAR}, including a held-out test set of 101 shots. The dataset employed in this work encompasses a diverse range of multimodal signals. While a comprehensive inventory of all variables is provided in \cref{input_data} of the Appendix, the primary signal categories include:
1)~\textbf{Plasma control inputs}, comprising actuator commands and plasma shape parameters prescribed by the plasma control system; 2)~\textbf{Plasma performance metrics}, representing volume-averaged confinement and stability indicators; and 3)~\textbf{MC measurements}, processed into cross-power and cross-phase spectrograms to capture magnetic fluctuation dynamics.

Plasma performance is characterised by two distinct scalar parameters. Normalised pressure, $\beta_N$, is a quantitative parameter of plasma stability that represents the global ratio of plasma pressure to magnetic pressure. A higher $\beta_N$ generally corresponds to greater fusion power, making its maximisation a central goal in fusion research. The confinement enhancement factor, $H_{89}$, is a quantitative indicator of plasma confinement performance. A larger $H_{89}$ implies that the plasma retains global energy more efficiently; therefore, both $\beta_N$ and $H_{89}$ should be maintained at high values. Notably, with the exception of the MC measurements, all other variables utilised here are scalar (0D) quantities.

\subsection{Mirnov Coil (MC)}
\label{Mirnov Coil (MC)}

The Mirnov coil (MC) is a fundamental diagnostic in tokamaks, providing time-series signals that measure magnetic field fluctuations ($\mathrm{d}B/\mathrm{d}t$) induced by MagnetoHydroDynamics (MHD) instabilities, which are the primary threats to plasma stability. These sensors, installed inside the vacuum vessel, detect fluctuations via Faraday’s law of induction. In KSTAR, MC signals are routinely available at a sampling rate of 2 MHz \cite{Bak2001RSI}, ensuring sufficient bandwidth to capture fast-evolving plasma dynamics.

To analyse the spatiotemporal characteristics of instabilities, we employ cross-spectral analysis between distinct MC channels \cite{choi_2025_RMPP}. Specifically, signal pairs undergo Fourier transformation to yield cross-power spectrograms, which quantify the intensity of shared fluctuations, and cross-phase spectrograms, which measure the phase delay between probe signals. These features are visualised in the time-frequency domain, enabling precise monitoring of the onset, frequency, and evolution of MHD instabilities.

\subsection{Data Preprocessing}
\label{Data Preprocessing}

Each shot is discretised into time slices with an interval of $\Delta t = 50\,\mathrm{ms}$, aligned with the temporal resolution of the magnetic EFIT \cite{Lao_1985} used for plasma shape parameters and performance metrics. To capture rapid modulations in KSTAR's external heating sources, we sampled heating power and positioning signals at a finer resolution of 5 ms. Thus, ten sequential scalar values are concatenated into the input vector for each global time step, $\Delta t$.

For MC inputs, we process signals from two representative channels via Fast Fourier Transform ($\mathrm{NFFT}=512, \mathrm{NSMTH}=3$) to yield cross-power and cross-phase spectrograms. Since magnetic fluctuations evolve on a faster timescale than the $\Delta t$, we stack 48 sequential spectral frames ($\approx 1$ ms resolution) for each 50 ms step, ensuring that high-frequency instability dynamics are preserved.

\subsubsection{MC Spectrogram Tokenisation}

\begin{figure}[ht]
  \vskip 0.2in
  \begin{center}
    \centerline{\includegraphics[width=\columnwidth]{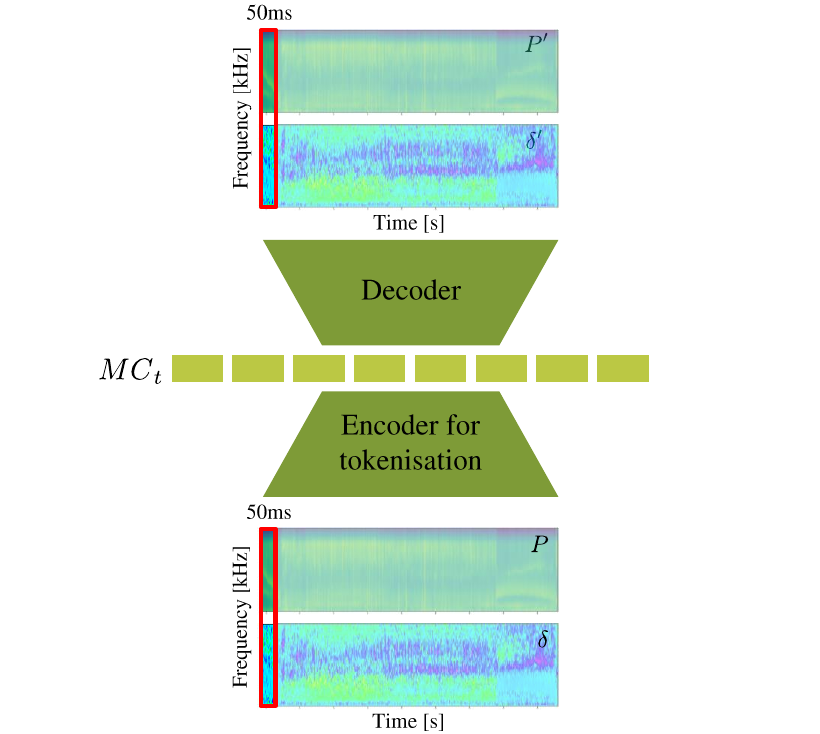}}
    \caption{\
    \textbf{VQ-VAE tokenisation process.} The VQ-VAE tokenises the input MC cross-power $P$ and cross-phase $\delta$ spectrograms - represented in $\mathbb{R}^{F \times T \times 2}$ (where $F$, $T$ denote frequency and time dimensions) -- into a discrete latent representation $MC_{t} \in \{1, \dots, V_o\}^{d_{MC}}$. Here $V_o$ represents the vocabulary size of the codebook, and $d_{MC}$ denotes the length of the token sequence.
    }
    \label{VQ-VAE}
  \end{center}
\end{figure}

The MC spectrograms evolve nonlinearly over time, and directly formulating their prediction as a regression problem often yields blurry outputs. To address this, we reframe the continuous spectrogram generation as a classification task over a learned discrete codebook. We employ a VQ-VAE \cite{VQ_VAE} to perform vector quantisation, mapping the continuous MC features to discrete latent indices. As illustrated in \cref{VQ-VAE}, input spectrograms processed at the global interval $\Delta t$ are encoded into a sequence of codebook indices $MC_t$, which are then reconstructed by the decoder during training.

\subsubsection{Scalar Values Tokenisation}

All scalar variables, excluding the MC spectrograms, undergo linear quantisation, mapping continuous values to discrete indices based on dataset-wide min-max statistics. We set the vocabulary size to $V_o=512$, which ensures high resolution; for example, the toroidal magnetic field in plasma control inputs is quantised with a step size of $\approx 4.0\times10^{-3}$. Following discretisation, the variables are concatenated into two distinct vectors: the \textit{plasma control token} $c_{t} \in \{1, \dots, V_o\}^{d_c}$, where $d_c=121$ represents the aggregated actuator and shape parameters, and the \textit{plasma performance token} $p_{t} \in \{1, \dots, V_o\}^{d_p}$, where $d_p=2$ corresponds to the plasma performance metrics.

\subsection{Model Architecture of PanoMHD}
\label{Model Architecture of PanoMHD}

\begin{figure}[ht]
  \vskip 0.2in
  \begin{center}
    \centerline{\includegraphics[width=\columnwidth]{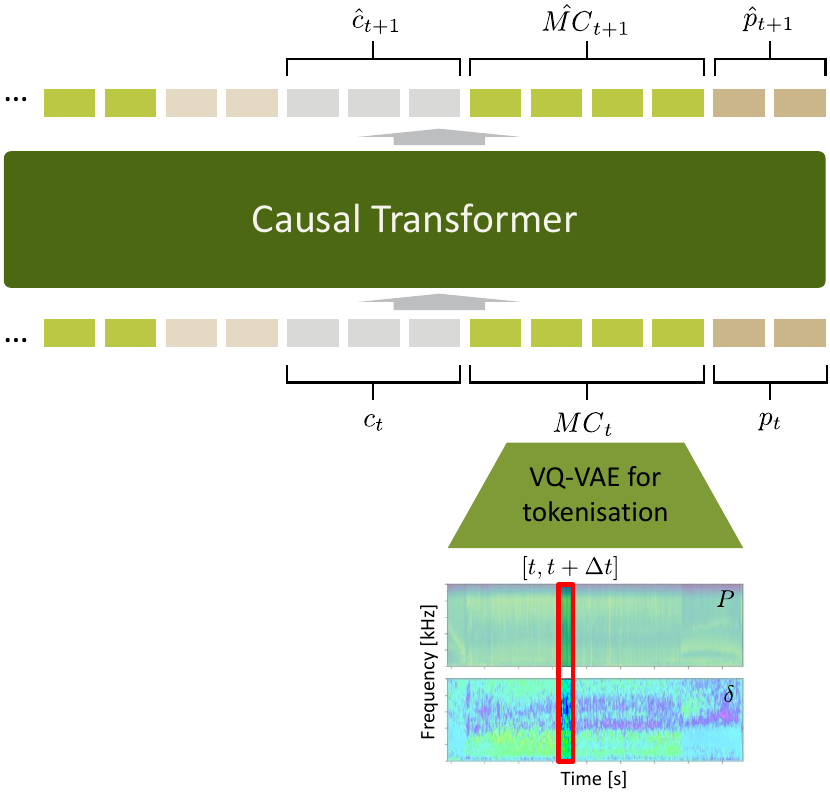}}
    \caption{
      \textbf{Overview of PanoMHD.} A causal Transformer predicts the next-step latent MC tokens and plasma performance metrics. The unified input sequence at each step comprises the plasma control $c_{t}$, latent MC $MC_t$, and plasma performance $p_t$. The respective token length ($d_c$, $d_{MC}$, and $d_p$) sum to a total sequence length of $d_{total}=219$. To predict the target states at $t+1$, the model processes a sequential context window of length $L=10$.
    }
    \label{PanoMHD}
  \end{center}
\end{figure}

Leveraging the discretised token sequences, we employ a decoder-only transformer \cite{attention} to model plasma dynamics. This approach is inspired by recent research showing the transformer architecture designed for LLMs can serve as an effective backbone for complex multimodal agents such as 3D game modelling \cite{AKanervisto_Nature_2025, genshin}. Motivated by this effectiveness, we adopted the GPT-2 architecture \cite{radford2019language} for PanoMHD, training from scratch to accommodate the distinct statistical properties of fusion plasma data, which diverge significantly from natural language corpora.

The model design illustrated in \cref{PanoMHD} shows the data processing flow for an arbitrary shot at time $t$. The input sequence is concatenated in the order of $c_{t}$, $MC_t$, and $p_t$. The model is conditioned on both the current plasma state ($MC_t, p_t$) and the plasma control ($c_{t}$) to predict the resulting latent magnetic fluctuations $\hat{MC}_{t+1}$ and plasma performance $\hat{p}_{t+1}$. 

\subsection{Evaluation Metrics for PanoMHD}
\label{evaluation}

PanoMHD is a multimodal framework, simultaneously predicting scalar plasma performance parameters of $\beta_N$ and $H_{89}$ as well as MC cross-power and cross-phase spectrograms. Furthermore, by integrating with Optimal Autoencoder-based State Identification System (OASIS) \cite{HJNoh_EPS_2025}, the framework extends to the downstream task of classification of the plasma confinement state of L-mode to H-mode (L/H) transition \cite{Wagner_PRL_1982}. OASIS is a deep learning-based L/H mode classifier validated on KSTAR that operates solely on MC spectrograms. In tokamaks, fusion plasmas transit from L-mode to H-mode when external heating exceeds a specific threshold \cite{FWagner_PRL_1984}. This transition is a central topic in fusion research \cite{Wagner_2007}, as H-mode typically achieves an energy confinement performance more than double that of L-mode plasmas \cite{MKeilhacker_1987}.

Evaluating AI models on scientific data can be challenging, particularly in fusion domain where there are no standardised or open benchmark datasets. For an AI model to be practically useful in a scientific domain, it must be assessed not only by statistical accuracy but also through quantitative metrics aligned with domain knowledge. To this end, we employ three complementary metrics for evaluating PanoMHD: $R^2$, PSNR and OASIS accuracy.

\subsubsection{Evaluating Plasma Performance Prediction}

The state of fusion plasma is governed by various physical factors, including MHD instabilities, and its performance is parameterised by $\beta_N$ and $H_{89}$. As MHD instabilities are captured by the MC signals and subsequently affect plasma performance, the regression performance on these performance metrics serves as an adequate indicator of whether the model has learned the latent causal relationship between instabilities and global plasma performance.

Notably, tearing instabilities -- MHD instabilities that break and reconnect magnetic field lines -- generally lead to sudden performance degradation, whereas the L/H mode transition triggers a rapid enhancement in confinement. Consequently, evaluating the model requires testing its ability to predict these abrupt shifts governed by nonlinear dynamics. To this end, we utilise the \textbf{Coefficient of Determination ($R^2$)}. As $\beta_N$ and $H_{89}$ are continuous scalar variables, $R^2$ provides a measure of how well the model accounts for the variance driven by these dynamical events.

\subsubsection{Evaluating Mirnov Coil Spectrogram Prediction}

\textbf{PSNR (Peak Signal-to-Noise Ratio)} is conventionally used to quantify the similarity between generated images and their ground truth counterparts. Given that PanoMHD predicts MC spectrograms as 2D time–frequency representations, PSNR serves as an appropriate measure to assess the similarity of the spectrograms. Because the numerical ranges of the MC spectrograms differ from standard 8-bit images, all data are linearly rescaled to the range $[0, 255]$ using global min–max statistics before computing PSNR.

As this study represents the pioneering effort to predict MC spectrograms for tokamak plasmas, there are no established baselines or benchmarks available for a direct quantitative comparison of PSNR. Furthermore, due to the lack of standardised metrics for evaluating the fidelity of generated MHD instability patterns, we complement the PSNR assessment with a qualitative analysis of physical signatures. Specifically, we inspect the predicted spectrograms for distinct visual features corresponding to major instabilities: Edge-Localised Modes (ELMs), which appear as broadband vertical stripes, and tearing instabilities, manifested as coherent horizontal streaks.

\subsubsection{Evaluating Downstream Task: L/H Mode Classification}

To evaluate the physical fidelity of our predictions beyond pixel-level metrics like PSNR, we employ the OASIS framework as a domain-specific evaluator. By applying OASIS to the generated MC spectrograms, we assess whether PanoMHD reconstructs distinct physical semantics rather than merely mimicking visual textures. High classification accuracy here is crucial; it demonstrates that the model has successfully encoded the latent physical coupling between magnetic fluctuations and the global confinement state.

The OASIS accuracy is computed by comparing the sequence of predicted L/H modes derived from PanoMHD’s output against the classification results from ground truth MC signals. In addition to this quantitative metric, we further validate the physical legitimacy of our model through a qualitative comparison with independent $D_\alpha$ emission measurements, which are the robust indicator to distinguish L/H modes. A supplementary explanation of $D_\alpha$ diagnostics is provided in the Appendix \cref{Dalpha}.

By jointly employing PSNR and OASIS accuracy, and independent physical validation, we evaluate PanoMHD from both statistical and physical perspectives, enabling a comprehensive assessment of the model’s predictive capability.

\section{Results}

\begin{table*}[t]
  \caption{\textbf{Comparison with baselines.} We report $R^2$ for scalar regression and accuracy for classification. 'N/A' indicates the model is not designed for that task.}
  \label{sota}
  \begin{center}
    \begin{small}
      \begin{sc}
        \begin{tabular}{lcccccr}
          \toprule
          Method & \makecell{$\beta_N$ \\ ($R^2$)} & \makecell{$H_{89}$ \\ ($R^2$)} & \makecell{Cross-power \\ (PSNR)} & \makecell{Cross-phase \\ (PSNR)} & \makecell{L/H mode Classification \\ (Accuracy)} \\
          \midrule
          Tearing prediction \cite{Seo2023IJCNNTearing}    & 0.957 & N/A & N/A & N/A & N/A \\
          L/H classification \cite{SHIN2020111634}        & N/A & N/A & N/A & N/A & 0.945 \\
          \textbf{PanoMHD (Ours)}  & \textbf{0.987} & 0.956 & 30.1 & 23.0 & \textbf{0.973} \\
          \bottomrule
        \end{tabular}
      \end{sc}
    \end{small}
  \end{center}
  \vskip -0.1in
\end{table*}

This section presents the evaluation of PanoMHD on a held-out test dataset of 101 KSTAR shots. We demonstrate that PanoMHD not only achieves SOTA performance on the established fusion tasks of future $\beta_N$ prediction and L/H mode classification, but also pioneers the prediction of high-dimensional MC spectrograms. A comprehensive comparison against existing baselines is provided in \cref{sota}. We first determine the optimal model configuration based on five key metrics: the $R^2$ for plasma performance parameters ($\beta_N$ and $H_{89}$), PSNR for MC cross-power and cross-phase spectrograms, and OASIS accuracy. Subsequently, we provide a qualitative analysis of the model’s predictive capabilities across representative KSTAR discharge scenarios.

\subsection{Performance of PanoMHD}

\begin{figure}[ht]
  \vskip 0.2in
  \begin{center}
    \centerline{\includegraphics[width=\columnwidth]{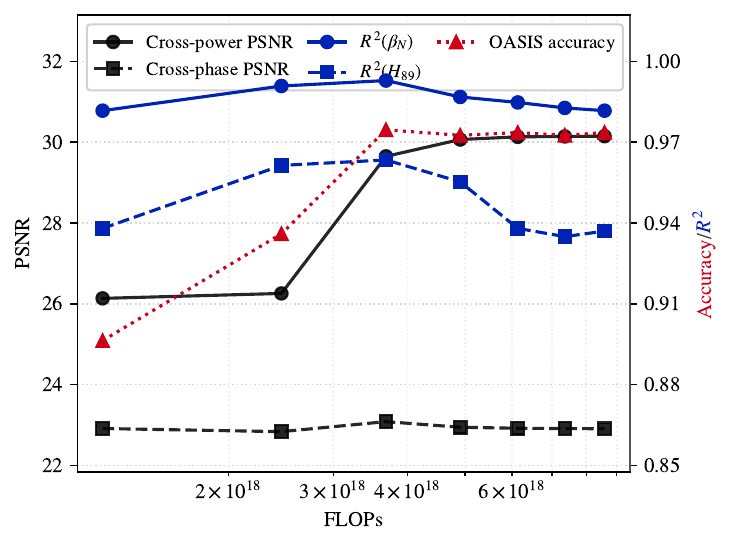}}
    \caption{
      \textbf{Evolution of PanoMHD evaluation metrics with respect to compute budget (FLOPs).} The plot tracks the test set performance for plasma parameters ($R^2$ of $\beta_N, H_{89}$), PSNR of MC cross-power, MC cross-phase, and L/H mode classification (OASIS accuracy). 
    }
    \label{FLOPs}
  \end{center}
\end{figure}

To identify the optimal model configuration, we analysed the trajectory of all five evaluation metrics against the cumulative floating-point operations (FLOPs) expended during training as illustrated in \cref{FLOPs}.

The learning dynamics reveal distinct convergence behaviours across modalities. Specifically, the scalar plasma parameters ($\beta_N, H_{89}$) and the cross-phase PSNR show rapid convergence in the early stages of training ($<2.5\times10^{18}$ FLOPs). This can be attributed to the simpler optimisation landscape of low-dimensional scalar performance tokens ($d_p=2$) compared to the high-dimensional MC spectrogram tokens ($d_{MC}=96$) representing 2D time–frequency signals. Notably, as MC cross-power prediction improves within the range of $(2.5\text{--}3.7)\times10^{18}$ FLOPs, we observe a slight but continuous enhancement in the scalar metrics. This suggests that the model progressively learns the latent correlations between magnetic fluctuations and global plasma performance, refining its scalar predictions based on better-understood spectral features.

In contrast, the cross-phase PSNR remains relatively static throughout training and is consistently lower than that of the cross-power. This disparity arises from the intermittent nature of MHD instabilities within the signal. Coherent phase differences between probe signals emerge at certain frequencies during MHD instabilities. In the absence of such instabilities, the phase difference becomes noise-dominated. Since PSNR averages pixel-level differences across the entire spectrogram, it is penalised by the mismatch in this background noise, even if the model accurately predicts the sparse but physically critical instability signatures. However, this lower PSNR does not imply a deficiency in model capability. The OASIS accuracy, which relies on both cross-power and cross-phase inputs for L/H mode classification, shows a steady improvement, ultimately nearing 1.0. This high classification performance confirms that PanoMHD successfully captures the essential physical semantics of both the cross-phase as well as the cross-power.

The performance reaches its peak at approximately $4.9\times10^{18}$ FLOPs. Beyond this point, we observe a performance degradation, indicating the onset of overfitting to the training data. Consequently, we established this peak as the early stopping criterion. The final PanoMHD model achieves $R^2(\beta_N)=0.987$, $R^2(H_{89})=0.956$, a cross-power PSNR of 30.1 dB, a cross-phase PSNR of 23.0 dB, and an OASIS accuracy of 97.3 \%.

\subsubsection{Prediction of MHD Instabilities}
\label{sec:instability_prediction}

\begin{figure}[ht]
  \vskip 0.2in
  \begin{center}
    \centerline{\includegraphics[width=\columnwidth]{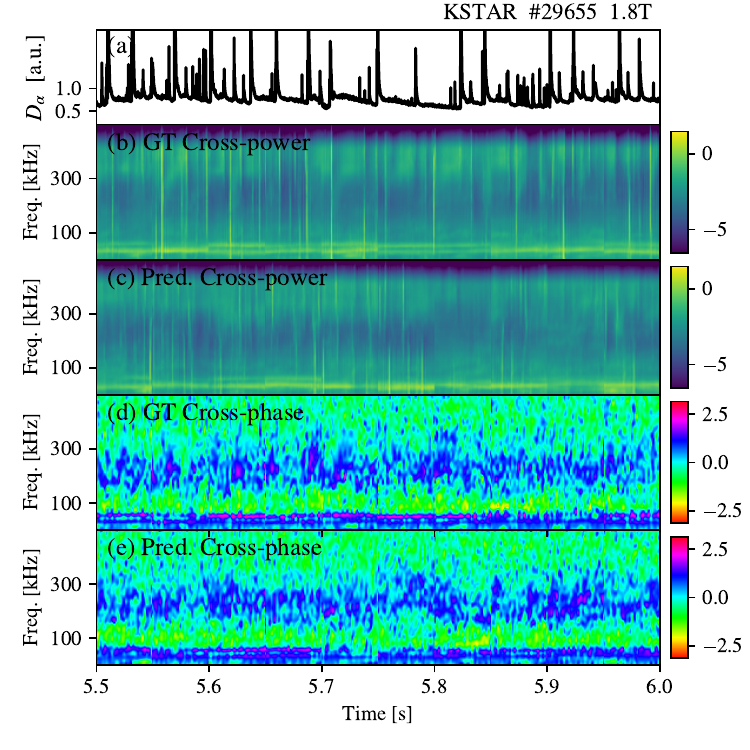}}
    \caption{
      \textbf{Zoomed-in analysis of ELM dynamics in KSTAR discharge \#29655.}
      The temporal window is restricted to $5.5\,\mathrm{s}$--$6.0\,\mathrm{s}$ to highlight the H-mode phase.
      (a) The $D_\alpha$ emission signal exhibits characteristic periodic spikes indicative of ELMs.
      (b), (c) Comparison of MC cross-power spectrograms between Ground Truth (GT) and PanoMHD prediction. The intermittent ELM events manifest as broadband vertical impulsive structures in the frequency domain, which are reproduced in the prediction.
      (d), (e) Comparison of MC cross-phase spectrograms between GT and PanoMHD prediction.
    }
    \label{zoom_29655}
  \end{center}
\end{figure}

\begin{figure}[ht]
  \vskip 0.2in
  \begin{center}
    \centerline{\includegraphics[width=\columnwidth]{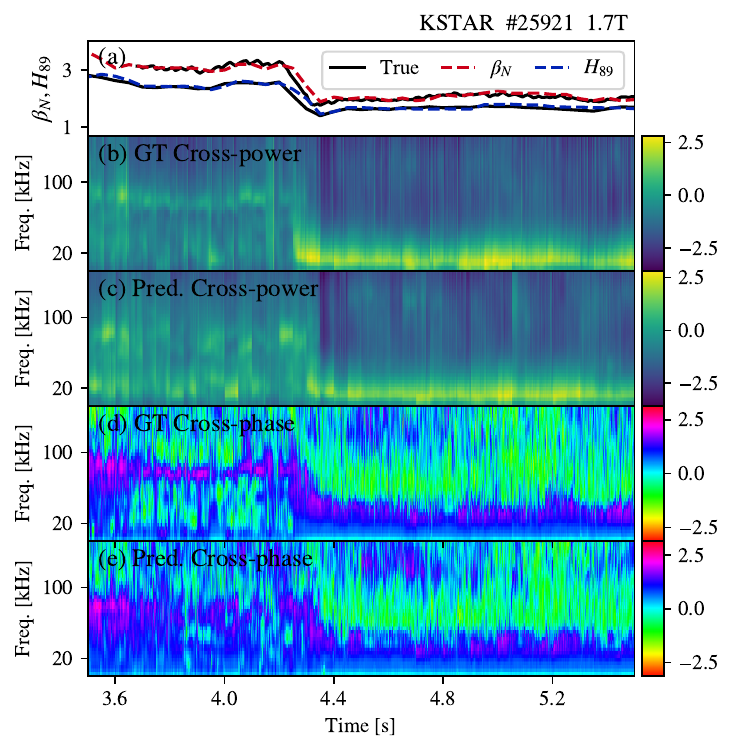}}
    \caption{
      \textbf{Zoomed-in analysis of tearing instabilities in KSTAR discharge \#25921.}
      The temporal window is restricted to $3.5\,\mathrm{s}$--$5.5\,\mathrm{s}$ to highlight the onset of tearing instabilities.
      (a) Comparison between true (solid black) and PanoMHD-predicted (dashed lines) plasma performance parameters ($\beta_N$ and $H_{89}$).
      (b-e) Comparison of MC cross-power and cross-phase spectrograms between GT and PanoMHD prediction.
    }
    \label{zoom_25921}
  \end{center}
\end{figure}

The fundamental capability of PanoMHD lies in its ability to predict the time–frequency evolution of magnetic fluctuations, which serve as direct signatures of MHD instability. We analyse the predicted spectrograms for two distinct instability scenarios: ELMs and tearing instabilities.

First, \cref{zoom_29655} illustrates H-mode dynamics dominated by ELMs. In the MC spectrogram, ELMs are characterised by broadband impulsive energy releases, manifesting as vertical stripes \cite{HZohm_1996_PPCF}. Notably, PanoMHD captures these events as distinct vertical features spanning the frequency range (\cref{zoom_29655}(b-c)), which align with the spikes in the independent $D_\alpha$ signal (\cref{zoom_29655}(a)).

Second, \cref{zoom_25921} highlights the onset of a tearing instability. Unlike the broadband nature of ELMs, tearing instabilities appear as coherent continuous signals with well-defined frequency (here, $\approx 20\,\mathrm{kHz}$ onset at 4.2 s), forming horizontal bands in the cross-power spectrogram. As shown in \cref{zoom_25921}(a), PanoMHD not only captures this instability but also predicts the consequent rapid degradation of plasma performance ($\beta_N$ and $H_{89}$).

\subsubsection{Prediction of Normalised Pressure $\beta_N$}

\begin{figure}[ht]
  \vskip 0.2in
  \begin{center}
    \centerline{\includegraphics[width=\columnwidth]{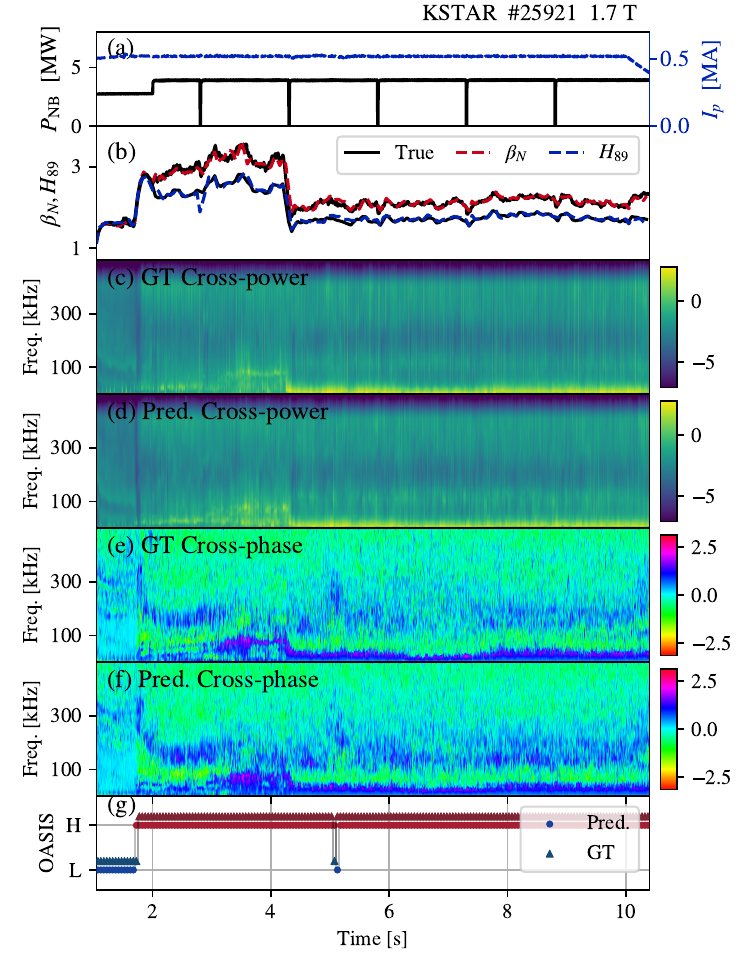}}
      \caption{
        \textbf{PanoMHD prediction on KSTAR discharge \#25921.} (a) Actuator inputs showing plasma current ($I_p$) and total NB heating power ($P_{\mathrm{NB}}$). (b) Comparison between true and predicted plasma performance parameters.
        (c-f) Comparison of MC cross-power and cross-phase spectrograms between GT and PanoMHD prediction.
        (g) Sequence of L/H mode classification derived from OASIS applied to both GT and Predicted MC spectrograms.
        }
    \label{shot25921}
  \end{center}
\end{figure}

Crucially, the accurate prediction of MHD instabilities translates directly into precise plasma performance prediction. We evaluate PanoMHD on the prediction of normalised pressure ($\beta_N$) by benchmarking it against a recent baseline \cite{Seo2023IJCNNTearing, Seo2024NatureRL}.

Tearing stability is governed by local pressure gradients at the specific radial location where the instability occurs \cite{KKim, mscha}. Consequently, the baseline relies on spatially resolved 1D measurements of electron density and temperature (via Thomson scattering diagnostics) to capture these local dependencies. However, obtaining such information in high-temperature tokamaks involves complex diagnostic processes that are often prone to measurement errors and latency.

In contrast, PanoMHD achieves superior performance ($R^2=0.987$ vs. $0.957$) utilising only fundamental 0D parameters and MC signals. This result suggests that, by scaling up model capacity with a multimodal causal Transformer on tokenised representations, PanoMHD effectively learns the causal dependency of plasma performance on MHD instabilities, thereby bypassing the need for expensive spatial diagnostics.

To visually corroborate these results, we examine the evolution of KSTAR shot \#25921 in \cref{shot25921}, which features the tearing instability discussed in \cref{sec:instability_prediction}. PanoMHD achieves a cross-power PSNR of 26.88 dB and a cross-phase PSNR of 20.76 dB for this discharge. Physically, the plasma maintains high performance ($\beta_N > 3.0$) until the instability triggers a rapid degradation, reducing $\beta_N$ to approximately 2.0. Crucially, PanoMHD leverages its multimodal architecture to capture this causal chain, correctly coupling the high-dimensional MC signature of the tearing mode (\cref{shot25921}(d)) with the resulting 0D performance degradation (\cref{shot25921}(b)).

In \cref{shot25921}(g), OASIS identifies the H-mode transition at 1.77 s (ground truth) and 1.72 s (prediction), achieving 0.984 accuracy for this shot. Given the resolution of OASIS and prediction step size ($\Delta t = 50\,\mathrm{ms}$), this single-step discrepancy indicates high temporal fidelity. Consistent with this transition, PanoMHD tracks the rapid increase in $\beta_N$ and $H_{89}$ immediately, as shown in \cref{shot25921}(b).

\subsubsection{Prediction of L/H mode Transition}

\begin{figure}[ht]
  \vskip 0.2in
  \begin{center}
    \centerline{\includegraphics[width=\columnwidth]{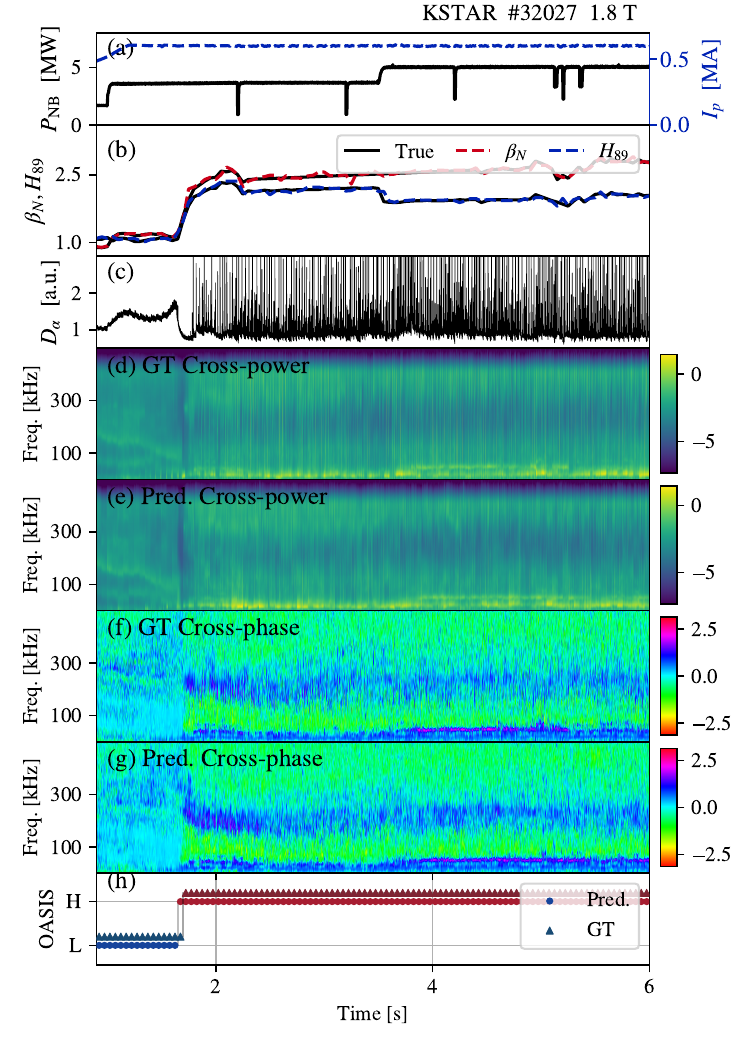}}
      \caption{
        \textbf{PanoMHD prediction on KSTAR discharge \#32027.} (a) Plasma control inputs. (b) Comparison of true and predicted plasma performance parameters. (c) The $D_\alpha$ emission signal. Note that $D_\alpha$ is an independent experimental measurement not used as input for PanoMHD; it serves here as a ground truth indicator for distinguishing L-mode and H-mode. (d-g) Visual comparison of GT and Predicted MC spectrograms. (h) OASIS L/H mode classification results.
        }
    \label{shot32027}
  \end{center}
\end{figure}

Beyond capturing MHD instabilities, PanoMHD serves as a foundation for the downstream task of L/H mode prediction. We benchmark against a baseline \cite{SHIN2020111634}, an LSTM-based model designed for automated plasma control applications, classifying L/H mode with 94.5\% accuracy.

The operational paradigms between the two approaches differ distinctly. The baseline determines the \textit{current} L/H mode using real-time measurements of electron density and $D_\alpha$, via a lightweight architecture. In contrast, our framework employs OASIS to infer \textit{future} transitions based on PanoMHD-predicted spectrograms. Despite the increased task difficulty, PanoMHD achieves 97.3\% accuracy, surpassing the baseline. This validates that predicted spectrograms are not merely visual approximations but encode the physical semantics required for mode classification.
 %It demonstrates that PanoMHD can extrapolate global confinement states (L/H mode) from local sensor predictions, capturing essential physics beyond simple instability signatures.

To visually corroborate this, we present KSTAR shot \#32027 in \cref{shot32027}, which features a clear L/H mode transition. Given that this transition is physically characterised by a visible drop in the $D_\alpha$ baseline followed by the emergence of ELMs, we utilise the $D_\alpha$ emission (\cref{shot32027}(c)) as an independent ground truth.

PanoMHD's multimodal predictions align well with this $D_\alpha$ signal. As confirmed by the drop in $D_\alpha$, the model predicts a simultaneous rise in plasma performance metrics ($\beta_N$ and $H_{89}$) in \cref{shot32027}(b), reflecting the improved confinement characteristic of the H-mode. Furthermore, the predicted MC spectrograms (\cref{shot32027}(e, g)) reproduce the onset of ELM following the transition. Quantitatively, OASIS identifies the transition time at 1.72 s (ground truth) and 1.67 s (prediction), yielding an accuracy of 0.995 for this discharge. The model achieves a cross-power PSNR of 25.14 dB and a cross-phase PSNR of 20.31 dB for this shot. Magnified comparisons of the transition dynamics are provided in \cref{zoom_25826} and \cref{zoom_32027} in the Appendix.

\section{Conclusion}
We introduced PanoMHD, a dynamic modelling framework that shifts the paradigm of tokamak instability prediction from isolated scalar indices to a comprehensive, multimodal representation integrating high-dimensional magnetic fluctuation spectrograms with global performance metrics.

By tokenising multimodal physical signals and training a causal Transformer, PanoMHD successfully captures the nonlinear evolution of fusion plasmas without complex diagnostics, making it viable solution for future commercial reactors. KSTAR evaluations demonstrate that PanoMHD (1) achieves SOTA performance in predicting plasma performance ($\beta_N$, $R^2=0.987$) and downstream plasma confinement mode classification (97.3\% accuracy), outperforming baselines; and (2) accurately captures MHD instability signatures, such as tearing instabilities and ELMs, within predicted magnetic fluctuations, validating physical consistency through coherent prediction of associated plasma performance changes.

A limitation of this study is its exclusive reliance on KSTAR data. However, as the first attempt to predict tokamak magnetic fluctuations via AI, this work establishes a critical baseline. Importantly, PanoMHD's methodology of modelling latent causal relationships using tokenised physical signals is generic. We thus envision this framework extending to diverse fusion devices, paving the way for multi-machine foundation models.

% \section*{Acknowledgements}
% The authors would like to express their deepest gratitude to KSTAR team. The authors thank Donghyuk Kim, Minsu Kim, and Hyunseok Lee for helpful discussions. This research was supported by the MSIT(Ministry of Science, ICT), Korea, under the Global Research Support Program in the Digital Field program(RS-2024-00436680) supervised by the IITP(Institute for Information & Communications Technology Planning & Evaluation). This project is supported by Microsoft Research Asia.

\section*{Impact Statement}

This work advances machine learning for scientific modelling by introducing a multimodal dynamics model that predicts magnetic fluctuations in tokamak fusion plasmas. The primary aim of PanoMHD is to support improved understanding of plasma behaviour. Accurate prediction of magnetic diagnostics may contribute to long term progress in plasma stability studies, potentially aiding the development of safer and more efficient fusion experiments. There are many potential societal consequences of our work, none of which we feel must be specifically highlighted here, as no concerns beyond those commonly associated with scientific machine learning research have been identified.


\bibliography{example_paper}
\bibliographystyle{icml2026}

%%%%%%%%%%%%%%%%%%%%%%%%%%%%%%%%%%%%%%%%%%%%%%%%%%%%%%%%%%%%%%%%%%%%%%%%%%%%%%%
%%%%%%%%%%%%%%%%%%%%%%%%%%%%%%%%%%%%%%%%%%%%%%%%%%%%%%%%%%%%%%%%%%%%%%%%%%%%%%%
% APPENDIX
%%%%%%%%%%%%%%%%%%%%%%%%%%%%%%%%%%%%%%%%%%%%%%%%%%%%%%%%%%%%%%%%%%%%%%%%%%%%%%%
%%%%%%%%%%%%%%%%%%%%%%%%%%%%%%%%%%%%%%%%%%%%%%%%%%%%%%%%%%%%%%%%%%%%%%%%%%%%%%%
\newpage
\appendix
%\onecolumn
\section{Appendix}

\begin{table*}[t]
  \caption{Data used for PanoMHD.}
  \label{input_data}
  \begin{center}
    \begin{small}
      \begin{sc}
        \begin{tabular}{lcccr}
          \toprule
          \textbf{Name} (inputs) & \textbf{Units} & \textbf{Type} & \textbf{Time}  \\
          \midrule
          Toroidal magnetic field       & $\mathrm{T}$ & Actuator & $t+\Delta t$ \\
          Plasma current                & $\mathrm{MA}$ & Actuator & $t+\Delta t$ \\
          Power of EC heating channel 2-5       & $\mathrm{MW}$ & Actuator & $t+\Delta t$ \\
          Resonant ($\mathrm{r}$, $\mathrm{z}$) position of EC channel 2-5 & $\mathrm{m}$ & Actuator & $t+\Delta t$ \\
          Power of NB1 heating channel A-C      & $\mathrm{MW}$ & Actuator & $t+\Delta t$ \\
          Power of NB2 heating channel A-C      & $\mathrm{MW}$ & Actuator & $t+\Delta t$ \\
          Fuelling GAS1-3, PVD          & $\mathrm{V}$ & Actuator & $t+\Delta t$ \\
          Minor radius                  & $\mathrm{m}$ & Plasma shape & $t+\Delta t$ \\
          Major radius                  & $\mathrm{m}$ & Plasma shape & $t+\Delta t$ \\
          Separatrix position ($\mathrm{dR_{sep}}$)  & $\mathrm{m}$ & Plasma shape & $t+\Delta t$ \\
          Diverted/limited configuration  & $\mathrm{-}$ & Plasma shape & $t+\Delta t$ \\
          Elongation  & $\mathrm{-}$ & Plasma shape & $t+\Delta t$ \\
          Upper triangularity  & $\mathrm{-}$ & Plasma shape & $t+\Delta t$ \\
          Lower triangularity  & $\mathrm{-}$ & Plasma shape & $t+\Delta t$ \\
          Normalised pressure, $\beta_N$  & $\mathrm{-}$ & Magnetic measurement & $t$ \\
          Confinement enhancement factor, $H_{89}$  & $\mathrm{-}$ & Magnetic measurement & $t$ \\
          Mirnov coil  & $\mathrm{-}$ & Mirnov measurement & $t$ \\
          \midrule
          \textbf{Name} (output) & \textbf{Units} & \textbf{Type} & \textbf{Time}  \\
          \midrule
          Normalised pressure, $\beta_N$  & $\mathrm{-}$ & Magnetic measurement & $t+\Delta t$ \\
          Confinement enhancement factor, $H_{89}$  &
          $\mathrm{-}$ & Magnetic measurement & $t+\Delta t$ \\
          Mirnov coil  & $\mathrm{-}$ & Mirnov measurement & $t+\Delta t$ \\
          \bottomrule
        \end{tabular}
      \end{sc}
    \end{small}
  \end{center}
  \vskip -0.1in
\end{table*}

\subsection{Detailed Description of Fusion Data Variables}
The dataset utilised in this work is summarised in \cref{input_data}. We use KSTAR experimental data of campaign years 2017--2022. For each shot in the campaign years, we exclude the early ramp up phase of plasma formation by selecting only time points with a plasma current above 100 kA. Shots with very short pulse durations ($t_{\mathrm{pulse}} < 1.7\mathrm{s}$) were removed and also, very long discharges ($t_{\mathrm{pulse}} > 20\mathrm{s}$) were excluded due to wall saturation effects. All data were retrieved from the KSTAR MDSplus repository, a standard software suite for data acquisition and storage widely adopted across the international fusion research community. Detailed descriptions of these variables are provided below to assist readers unfamiliar with tokamak physics.

\begin{itemize}
    \item \textbf{Actuators (Control Inputs):} These variables represent the operational commands applied to the tokamak to sustain and control the plasma.
    \begin{itemize}
        \item \textbf{Toroidal Magnetic Field ($B_T$) \& Plasma Current ($I_p$):} The fundamental electromagnetic constraints required to confine the high-temperature plasma within the vacuum vessel.
        \item \textbf{External Heating (EC \& NB):} Energy injection systems used to heat the plasma beyond its intrinsic ohmic heating. 
        \begin{itemize}
            \item \textbf{Electron Cyclotron (EC):} Heating via microwave injection, defined by the injected power (MW) and the resonant position (deposition location, $r, z$).
            \item \textbf{Neutral Beam (NB):} Heating via the injection of high-energy neutral particles, specified by the power (MW) of each beamline.
        \end{itemize}
        \item \textbf{Fuelling:} The voltage applied to piezoelectric valves to inject gas (typically deuterium) into the vacuum vessel, thereby controlling the plasma density.
    \end{itemize}

    \item \textbf{Plasma Shape (Geometry):} These parameters define the 2D cross-sectional shape of the plasma boundary, which critically influences stability and confinement.
    \begin{itemize}
        \item \textbf{Major Radius ($R$) \& Minor Radius ($a$):} Geometric parameters defining the size and position of the plasma torus.
        \item \textbf{Elongation ($\kappa$) \& Triangularity ($\delta$):} Parameters describing the deviation of the plasma cross-section from a perfect circle. Modern tokamaks typically shape plasmas into a "D-shape" (high $\kappa$ and $\delta$) to improve confinement properties.
        \item \textbf{Separatrix Distance ($dR_{sep}$) \& Configuration:} $dR_{sep}$ measures the distance between magnetic field lines at specific locations. It determines whether the plasma is in a \textbf{diverted} configuration (possessing a magnetic X-point, standard for high-performance H-mode) or a \textbf{limited} configuration (in direct contact with the limiter wall).
    \end{itemize}

    \item \textbf{Plasma Performance \& Measurements (State Outputs):} Key scalar and spectral indicators representing the physical state of the plasma.
    \begin{itemize}
        \item \textbf{Normalised Pressure ($\beta_N$):} The ratio of plasma pressure to magnetic pressure. A higher $\beta_N$ indicates higher fusion performance but often correlates with an increased risk of instabilities.
        \item \textbf{Confinement Enhancement Factor ($H_{89}$):} A normalised metric comparing the plasma's energy confinement time to the standard L-mode scaling law. Consequently, $H_{89} \approx 1$ corresponds to a typical L-mode plasma, whereas H-mode plasmas generally exhibit improved confinement with values of $H_{89} \approx 2$.
        \item \textbf{Mirnov Coils (MC):} High-frequency magnetic sensors that measure magnetic fluctuations.
    \end{itemize}
\end{itemize}

\subsubsection{Definition of $\beta_N$ and $H_{89}$}
To evaluate and compare plasma performance across different tokamak devices regardless of their varying sizes and operational capabilities (e.g., magnetic field strength and plasma current), we utilise two fundamental dimensionless parameters: the normalised beta ($\beta_N$) and the confinement enhancement factor ($H_{89}$). These normalised metrics serve as standard figures of merit for quantifying tokamak plasma performance and confinement quality.

The normalised pressure is defined as
\[
\beta_N = \beta \, \frac{a B_t}{I_p},
\]
where
\[
 \beta = \frac{\langle p \rangle}{B^2 / (2\mu_0)}.
\]
Here, $\langle p \rangle$ is the mean plasma pressure, $B$ is the total magnetic field magnitude, $B_t$ is the toroidal magnetic field, $\mu_0 = 4\pi \times 10^{-7}\,\mathrm{H/m}$ is the vacuum permeability, $a$ is the minor radius, and $I_p$ is the plasma current.
This dimensionless parameter represents the ratio of plasma pressure to magnetic pressure and quantifies how close the plasma operates to high pressure stability limits.

The confinement enhancement factor is defined as 
\[
H_{89} = \frac{\tau_E^{\mathrm{measured}}}{\tau_E^{\mathrm{ITER89P}}},
\]
where $\tau_E^{\mathrm{measured}}$ denotes the experimentally obtained energy confinement time (representing the characteristic exponential decay time of the stored plasma energy) and $\tau_E^{\mathrm{ITER89P}}$ is the predicted L mode confinement time given by the ITER89P scaling law across multiple tokamaks:
\[
\tau_E^{\mathrm{ITER89P}} = 0.048\, M^{0.5} I_p^{0.85} R^{1.2} a^{0.3} \kappa^{0.5} n^{0.1} B_t^{0.2} P_T^{-0.5},
\]
where $M$ is the average isotopic mass, $R$ is the major radius, $\kappa$ is the plasma elongation, $n$ is the line averaged density, and $P_T$ is the total heating power \cite{Yushmanov_1990}. Thus, $H_{89}$ represents the ratio of the plasma’s confinement performance to the empirical reference value $\tau_E^{\mathrm{ITER89P}}$.

\subsubsection{Cross-spectral Analysis of MC}

For an MC signal measured in the i-th channel, $\mathrm{d}B_i(t)/\mathrm{d}t$, let $X_i(f)$ denote its complex Fourier transform. The cross power is defined as $P = |X_i X_j^\dagger|$, and the cross phase is defined as $ \delta = \tan^{-1}\left(\frac{\mathrm{Im}(X_i X_j^\dagger)}{\mathrm{Re}(X_i X_j^\dagger)}\right). $ Here, $X_j^\dagger$ is the complex conjugate of the j-th channel’s Fourier transform, while $\mathrm{Re}$ and $\mathrm{Im}$ denote the real and imaginary parts, respectively \cite{kim_mhd_1999}. Because the cross-phase is expressed in radians, its values naturally lie within the interval $-\pi$ to $\pi$. 

\subsection{Physical Validation Signal: $D_\alpha$ Emission}
\label{appendix_d_alpha}

In this section, we provide a detailed description of the Deuterium-alpha ($D_\alpha$) emission signal, which is utilised in this study as an independent ground truth for physically validating the model's predictions, particularly for identifying L-mode to H-mode transitions.

\subsubsection{Description of $D_\alpha$ Emission Signal}
\label {Dalpha}

Although not utilised as an input feature for PanoMHD, the $D_\alpha$ signal is presented in our analysis as a critical, independent experimental validation metric. The $D_\alpha$ signal measures the intensity of visible light (at 656.3 nm) emitted from deuterium atoms undergoing the Balmer-alpha transition.

Physically, this emission occurs primarily at the cooler edge of the plasma and serves as a proxy for \textbf{particle recycling} -- the rate at which plasma particles escape confinement, strike the vessel wall, and re-enter as neutrals. The $D_\alpha$ intensity provides two distinct signatures for identifying confinement state and ELMs:

\begin{itemize}
    \item \textbf{L-H Transition (Baseline Drop):} When the plasma transitions from Low-confinement (L-mode) to High-confinement (H-mode), a transport barrier forms at the plasma edge. This barrier drastically reduces the flux of particles escaping to the wall, leading to a sudden decrease in recycling. Consequently, the onset of H-mode is marked by a sharp, characteristic drop in the $D_\alpha$ baseline intensity.
    
    \item \textbf{Edge-Localised Modes (ELMs):} Following the transition to H-mode, the steep pressure gradient at the edge can trigger quasi-periodic MHD instabilities known as ELMs. These events expel bursts of energy and particles to the wall, manifesting as rapid, high-amplitude spikes superimposed on the low H-mode baseline. The management of these events is critical in fusion reactors; the expulsion of high-temperature particles and stored energy can cause damage to plasma-facing components \cite{HZohm_1996_PPCF}. Consequently, reliable ELM prediction and active suppression are essential research objectives \cite{SHIN2018341, Shin_2022}.
\end{itemize}

\begin{figure}[ht]
  \vskip 0.2in
  \begin{center}
    \centerline{\includegraphics[width=\columnwidth]{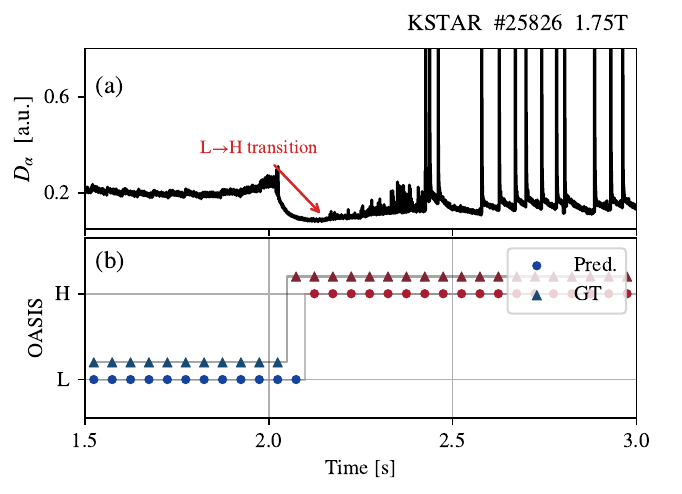}}
    \caption{
      \textbf{Zoomed analysis of L/H transition in KSTAR discharge \#25826.} 
      The temporal window is restricted to $1.5\,\mathrm{s}$--$3.0\,\mathrm{s}$ to highlight the transition dynamics. 
      (a) The $D_\alpha$ emission signal. The specific moment of the L/H transition is identified by the characteristic drop in the signal baseline, as indicated by the red arrow. 
      (b) Comparison of L/H mode classification results generated by OASIS using GT and PanoMHD-predicted MC spectrograms.
    }
    \label{zoom_25826}
  \end{center}
\end{figure}

\begin{figure}[ht]
  \vskip 0.2in
  \begin{center}
    \centerline{\includegraphics[width=\columnwidth]{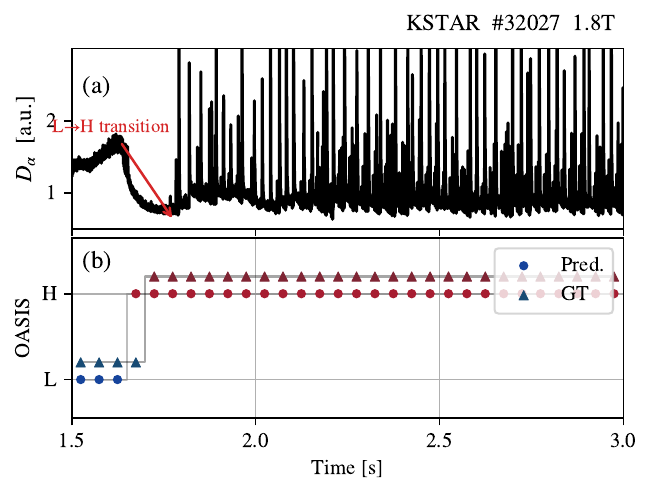}}
    \caption{
      \textbf{Zoomed analysis of KSTAR discharge \#32027.} 
      The plotting format is identical to \cref{zoom_25826}. 
      (a) The $D_\alpha$ signal with a red arrow indicating the L/H transition time. 
      (b) Comparison of OASIS L/H classification results between GT and PanoMHD predictions.
    }
    \label{zoom_32027}
  \end{center}
\end{figure}

\subsubsection{Validation of L/H Mode Transitions via $D_\alpha$ Signals}
\label{detailed_case_studies}

The $D_\alpha$ signal serves as the ground truth for determining the exact timing of confinement mode transitions. Referring to \cref{zoom_25826}(a) and \cref{zoom_32027}(a), the distinctive drop in the $D_\alpha$ baseline clearly indicates the onset of the H-mode phase, followed by ELM spikes. Comparing these spectral signatures with the predictions in panel (b) of both figures reveals that PanoMHD accurately pinpoints the transition timing. Crucially, this physical consistency with an independent diagnostic provides robust cross-validation, reinforcing the high classification accuracy observed in the OASIS framework. This alignment confirms that the model successfully captures global confinement state dynamics without direct access to the $D_\alpha$ diagnostic.

\subsection{Implementation Details and Additional Analysis}

In this section, we provide detailed training configurations to ensure reproducibility, followed by an analysis of the VQ-VAE's reconstruction quality and an ablation study on model scaling.

\subsubsection{Training Configuration of PanoMHD}
We train the PanoMHD model using the AdamW optimiser \cite{AdamW} with a constant learning rate of $8.0\times10^{-4}$ preceded by a linear warm-up. Hyperparameters were set to $\beta_1 = 0.9$, $\beta_2 = 0.95$, and $\epsilon = 1.0\times10^{-4}$. The model was trained with a batch size of 4. Training on four NVIDIA A100 GPUs required approximately 5.5 hours.

\subsubsection{Training Configuration of VQ-VAE}
The encoder consists of three convolutional stages, each implemented as two 3×3 Conv–Batch Normalisation–ReLU blocks followed by a 2×2 max pool. The resulting latent feature map undergoes vector quantisation utilising a codebook of $V_o = 512$ embeddings, each with a dimensionality of $D = 128$. Subsequently, this quantised representation is reshaped into a discrete sequence comprising $d_{MC} = 96$ tokens. The decoder mirrors the encoder with three convolutional stages (each containing two 3×3 Conv–BN–ReLU blocks) and applies bilinear upsampling to recover the original 2D MC spectrogram shape.

For the VQ-VAE tokeniser, the input dimensions were set to $(F, T, C) = (128, 48, 2)$. Here, $F$ and $T$ represent the frequency (vertical axis) and time (horizontal axis) dimensions, respectively, while the channel dimension $C$ corresponds to the cross-power and cross-phase spectrograms. We employed the mean squared error (MSE) loss function and the Adam optimiser \cite{Adam} with a constant learning rate of $1.0\times10^{-3}$ ($\beta_1 = 0.9, \beta_2 = 0.999, \epsilon = 1.0\times10^{-8}$). To prevent overfitting, we implemented early stopping with a patience of 50 epochs. Training was performed on a single NVIDIA RTX 3090 GPU and took approximately 16 hours.

\subsubsection{Analysis of VQ-VAE Reconstruction Quality}
\label{vqvae_analysis}

Since PanoMHD operates on tokens derived from the VQ-VAE, the reconstruction quality of the VQ-VAE establishes the theoretical upper bound for the model's predictive performance. To validate the efficacy of our tokenisation strategy, we evaluated the reconstruction capability of the VQ-VAE on the test dataset.

\begin{table}[t]
  \caption{Reconstruction performance of VQ-VAE in PSNR (dB).}
  \label{vqvae_table}
  \begin{center}
    \begin{small}
      \begin{sc}
        \begin{tabular}{lccr}
          \toprule
            Metric & Cross-power & Cross-phase \\
          \midrule
            Reconstruction & 31.6 & 24.1 \\
          \bottomrule
        \end{tabular}
      \end{sc}
    \end{small}
  \end{center}
  \vskip -0.1in
\end{table}

Table \ref{vqvae_table} presents the average PSNR between the original ground truth spectrograms and those reconstructed from the discrete tokens. The VQ-VAE achieves a reconstruction PSNR of 31.6 dB for cross-power and 24.1 dB for cross-phase, which are higher than the predictive performance of PanoMHD (30.1 dB and 23.0 dB, respectively).

\subsubsection{Ablation Study on PanoMHD}

\begin{table*}[t]
  \caption{\textbf{Ablation study on model scaling.} Comparison of predictive performance between 457M and 205M parameter variants of PanoMHD. The evaluation spans scalar plasma parameters ($R^2$), MC spectrograms (PSNR [dB]), and downstream L/H mode classification (OASIS accuracy).}
  \label{ablation}
  \begin{center}
    \begin{small}
      \begin{sc}
        \begin{tabular}{lccccccccr}
          \toprule
           Model Variant & Params & \makecell{$R^2(\beta_N)$ \\ ($\uparrow$)} & \makecell{$R^2(H_{89})$ \\ ($\uparrow$)} & \makecell{Cross-power \\ PSNR ($\uparrow$)} & \makecell{Cross-phase \\ PSNR ($\uparrow$)} & \makecell{OASIS \\ accuracy ($\uparrow$)}  \\
          \midrule
          457M PanoMHD & 457M & \textbf{0.987} & \textbf{0.956}  & \textbf{30.1} & \textbf{23.0} & \textbf{0.973}  \\
          205M PanoMHD & 205M  & 0.977 & 0.924 & 27.5 & 22.5 & 0.958  \\
          \bottomrule
        \end{tabular}
      \end{sc}
    \end{small}
  \end{center}
  \vskip -0.1in
\end{table*}

We conducted an ablation study to examine the impact of model scaling on predictive performance. We evaluated two model variants with different parameter sizes:
\begin{itemize}
\item \textbf{205M PanoMHD}: 12 layers, 12 heads, hidden size 768
\item \textbf{457M PanoMHD}: 36 layers, 16 heads, hidden size 1024
\end{itemize}
A comparison between the two configurations in \cref{ablation} demonstrates that the larger 457M model achieves superior performance across all metrics. Consequently, we adopted the 457M model as our final configuration.

%%%%%%%%%%%%%%%%%%%%%%%%%%%%%%%%%%%%%%%%%%%%%%%%%%%%%%%%%%%%%%%%%%%%%%%%%%%%%%%
%%%%%%%%%%%%%%%%%%%%%%%%%%%%%%%%%%%%%%%%%%%%%%%%%%%%%%%%%%%%%%%%%%%%%%%%%%%%%%%

\end{document}